\documentclass[a4paper,11pt]{article}
\pdfoutput=1 

\usepackage{jheppub} 

\usepackage[T1]{fontenc} 
\usepackage{appendix}
\usepackage[dvipsnames]{xcolor}

\usepackage{mathtools,slashed, nccmath, graphicx}
\usepackage{tikz}
\usepackage{subcaption}
\usepackage{graphicx}
\usepackage{bm}
\makeatletter
\gdef\@fpheader{\normalfont DOI: 10.1103/PhysRevD.111.103031 }
\makeatother

\title{Axion-Like Particle Mediated Dark Matter and Neutron Star Properties in the QHD Model}

\author[a]{Tanech Klangburam}
\author[a]{Chakrit Pongkitivanichkul}

\affiliation[a]{Khon Kaen Particle Physics and Cosmology Theory Group (KKPaCT),\\ Department of Physics, Faculty of Science, Khon Kaen University, 123 Mitraphap Rd.,\\ Khon Kaen, 40002, Thailand}

\emailAdd{klangburam.t@gmail.com}
\emailAdd{chakpo@kku.ac.th}

\abstract{
We investigate the effects of the ALP-mediated dark matter (DM) model on neutron star properties using the Quantum Hadrodynamics model (QHD). Using the relativistic mean-field approximation with the QHD-ALP-DM framework, we compute the equation of state (EoS) of neutron stars. Based on our previous study, we find that typical ALP parameter values have no significant effect on the EoS. We then explore various parametrizations of this model by varying the DM Fermi momentum, $q_f$, and DM mass, $m_{\chi}$. Our results show that increasing $q_f$ or $m_{\chi}$ shifts the energy density to higher values while reducing the maximum mass, radius, and tidal deformability of neutron stars. Finally, comparison with observational constraints from gravitational wave events and pulsar measurements indicates that the allowed parameter space for this model is constrained to $q_f < 0.05$ GeV and $m_{\chi} < 1000$ GeV. As a result, our study highlights the importance of next-generation gamma-ray observatories, such as the Cherenkov Telescope Array (CTA), in probing the ALP-mediated DM model.

}

\begin{document} 
\maketitle
\flushbottom

\section{Introduction}

According to astrophysical and cosmological observations, the majority of the mass content of the universe is dominated by dark matter (DM) \cite{ParticleDataGroup:2024cfk,Bertone:2004pz,Planck:2018vyg,WMAP:2003elm,COBE:1992syq}. Despite extensive efforts, the fundamental nature of DM remains elusive. Several detection methods have been developed, ranging from direct searches using particle accelerators \cite{ATLAS:2024kpy,CMS:2024zqs} and nuclear recoil experiments \cite{Bernabei:2013xsa,XENON:2022ltv,DarkSide:2022knj,PandaX:2022xqx} to indirect astrophysical probes \cite{Fermi-LAT:2015att,Fermi-LAT:2017opo,HESS:2016pst}. Among these approaches, neutron stars provide a unique opportunity to explore the interplay between particle physics and astrophysics, particularly in the context of beyond standard model (BSM) physics \cite{Bertone:2007ae,Kouvaris:2010vv,Deliyergiyev:2019vti,Barbat:2024yvi}. 

DM can influence neutron stars in various ways, including modifying their mass-radius relation 
\cite{Blinnikov:1983gh,Khlopov:1989fj,deLavallaz:2010wp,Li:2012ii,Sandin:2008db,Leung:2011zz,Leung:2012vea,Xiang:2013xwa,Goldman:2013qla,Khlopov:2013ava,Mukhopadhyay:2015xhs,Tolos:2015qra,Rezaei:2016zje,Panotopoulos:2017idn,Gresham:2018rqo,Deliyergiyev:2019vti,DelPopolo:2019nng,Karkevandi:2021ygv,Guha:2021njn,Miao:2022rqj,Sen:2022pfr,Ferreira:2022fjo,Shakeri:2022dwg,Hippert:2022snq,Cassing:2022tnn,Zollner:2022dst,Zollner:2023myk,Cronin:2023xzc,Sagun:2023rzp,Liu:2023ecz,Bramante:2023djs,Giangrandi:2024qdb,Guha:2024pnn,Routaray:2022utr,Routaray:2023spb}, altering tidal deformability \cite{Nelson:2018xtr,Ellis:2018bkr,Ivanytskyi:2019wxd,Dengler:2021qcq,Sen:2021wev,Collier:2022cpr,Lenzi:2022ypb,Leung:2022wcf,Thakur:2023aqm,Mariani:2023wtv,Diedrichs:2023trk,Karkevandi:2024vov,Liu:2024rix}, changing cooling rates \cite{Kouvaris:2007ay,Bertone:2007ae,Kouvaris:2010vv,McCullough:2010ai,deLavallaz:2010wp,Sedrakian:2018kdm,Bhat:2019tnz,Ema:2024wqr}. Depending on the nature of the DM candidate, these effects can range from subtle shifts in the equation of state (EoS) properties to more extreme cases such as DM forming a distinct core within the neutron star or accumulating in a diffuse halo. The presence of DM inside neutron stars has been studied in various contexts, including asymmetric DM, bosonic condensates, fermionic DM, weakly interacting massive particles (WIMPs), Higgs portal, and vector mediator, each leading to different astrophysical signatures \cite{Bell:2013xk,Mukhopadhyay:2016dsg,Panotopoulos:2018ipq,Rutherford:2022xeb,Karkevandi:2024vov,Thakur:2023aqm,Mariani:2023wtv,Diedrichs:2023trk,Kouvaris:2007ay,Das:2018frc,Dutra:2022mxl,Lourenco:2022fmf,Flores:2024hts,Bertoni:2013bsa}. Among these candidates, axion-like particles (ALPs) which are light pseudoscalar particles arising from extensions of the standard model, have gained significant attention due to their strong theoretical motivation \cite{Sedrakian:2018kdm,Klangburam:2023vjv}.

Quantum Hadrodynamics (QHD) provides a powerful theoretical framework for modeling the nuclear interactions inside neutron stars based on relativistic mean-field (RMF) approaches. QHD describes nuclear matter interactions via mesonic fields, including the scalar meson (responsible for attraction) and the vector meson (responsible for repulsion) \cite{Walecka:1974qa,Serot:1997xg,Serot:2000pv}. The QHD framework successfully explains nuclear saturation properties and forms the foundation for many neutron star EoS models. By incorporating an ALP-mediated DM interaction into QHD, we can investigate how these additional interactions modify the nuclear EoS and affect observable neutron star properties.

In this work, we investigate the impact of an ALP-mediated DM interaction on neutron star properties within the QHD framework. We derive the modified EoS by including ALP-nucleon and ALP-DM couplings in the relativistic mean-field approximation. We then solve the Tolman-Oppenheimer-Volkoff (TOV) equations to obtain the mass-radius relation and analyze the effect of ALPs on neutron star tidal deformability, an important parameter constrained by gravitational wave observations. Our study aims to determine whether ALP-mediated interactions lead to deviations from standard neutron star models that could be detected by current or future astrophysical observations.

The remainder of this paper is organized as follows. In Section \ref{sect:2}, we present the theoretical model, including the QHD framework and the ALP-mediated interaction. Section \ref{sect:3} describes the formalism used to solve for neutron star structure and tidal deformability. In Section \ref{sect:4}, we discuss our results, including the impact of ALP parameters on the EoS, mass-radius relation, and observational constraints. Finally, in Section \ref{sect:5}, we summarize our findings and outline potential directions for future research.



\section{Model}
\label{sect:2}

In this model, we consider that DM is being trapped inside the neutron star. The nucleon-nucleon interaction can be described by the $\sigma-\omega$ model, which is the simplest QHD model \cite{Walecka:1974qa,Serot:1997xg}. In this simplest version, the nucleon interaction is mediated by scalar and vector mesons ($\sigma$ and $\omega$), in which the scalar meson is responsible for the attractive force and the vector meson is responsible for the repulsive force. However, the original QHD model does not accurately reproduce nuclear bulk properties, such as the compressibility \cite{Serot:1997xg}. More advanced versions of QHD improve upon this by including additional interactions, such as the exchange of the isovector $\rho$ meson between nucleons and nonlinear interaction terms. The Lagrangian in this model is given by
\begin{eqnarray}
    \mathcal{L}_{HAD} &=& \bar{\psi}\left[ \gamma^\mu \left(i\partial_\mu - g_\omega \omega_\mu - g_\rho \vec{\tau}\cdot\vec{b} \right) - (M + g_\sigma \sigma) \right]\psi \nonumber \\
    & & + \frac{1}{2} \left( \partial_\mu \sigma \partial^\mu \sigma - m_\sigma^2 \sigma^2 \right) - \frac{1}{3}A\sigma^3 - \frac{1}{4}B\sigma^4 \nonumber \\
    & & - \frac{1}{4}\Omega_{\mu\nu}\Omega^{\mu\nu} + \frac{1}{2}m_\omega^2 \omega_\mu \omega^\mu - \frac{1}{4}R_{\mu\nu}R^{\mu\nu} + \frac{1}{2}m_\rho^2 \vec{b}_\mu \vec{b}^\mu , \label{eq:LQHD}
\end{eqnarray}
where $\psi$ is the nucleon field, $\omega_\mu$ is the vector meson, $\sigma$ is the scalar meson, and $\vec{b}$ is the isovector meson. The nucleon mass is $M=939$ MeV, and $m_\sigma$, $m_\omega$, and $m_\rho$ are different meson masses. The couplings $g_\sigma$, $g_\omega$, and $g_\rho$ are scalar, vector, and isovector coupling constants, respectively. The field strength tensor of the vector and isovector meson are given by $\Omega_{\mu\nu} = \partial_\mu \omega_\nu - \partial_\nu \omega_\mu$ and $R_{\mu\nu} = \partial_\mu \vec{b}_\nu - \partial_\nu \vec{b}_\mu$.


For DM part, we consider the ALP-mediated DM model previously considered in our previous work \cite{Klangburam:2023vjv}. The SM is extended by a Dirac fermion, $\chi$, and a pseudoscalar ALP, $a$. We assume that DM interacts with the nucleon through the ALP mediator. The effective Lagrangian is given by
\begin{eqnarray}
    \mathcal{L}_{DM} &=& \bar{\chi} ( i\gamma^\mu\partial_\mu - m_\chi ) \chi + \frac{1}{2}\partial_\mu a \partial^\mu a - \frac{1}{2}m_a^2 a^2  \nonumber \\
    & & + \sum_f \frac{m_\chi}{f_a} C_\chi \bar{\chi} i \gamma^5 \chi a + \sum_f \frac{m_f}{f_a} C_f \bar{\psi} i \gamma^5 \psi a, \label{eq:LDM}
\end{eqnarray}
where $f$ is any SM fermion, $m_f$, $m_\chi$, and $m_a$ are the masses of fermion, DM, and ALP, respectively. We define the effective ALP couplings as
\begin{eqnarray}
    g_{aff} = \frac{m_f}{f_a}C_f\quad \text{and}\quad g_{a\chi\chi} = \frac{m_\chi}{f_a}C_\chi,
\end{eqnarray}
where the ALP-fermion couplings $C_f$ are assumed to be universal for any SM fermion, i.e., $C_f$ and $g_{aff} \propto m_f$. 
The various regimes of the ALP-mediated DM model have been studied in the literature, for example, freeze-in/out scenarios have been studied in \cite{Bharucha:2022lty,Ghosh:2023tyz,Dror:2023fyd,Armando:2023zwz,Allen:2024ndv}, while study with astrophysical observations has been studied in \cite{Klangburam:2023vjv,Yang:2024jtp}. In this work, we will focus on the case that DM interacts with nucleons in the neutron star. The total Lagrangian of QHD-ALP-DM system is 
\begin{eqnarray}
    \mathcal{L} = \mathcal{L}_{HAD} + \mathcal{L}_{DM}.
\end{eqnarray}

By using the relativistic mean-field approximation, the system is assumed to be uniform in the ground state. The fields, including the meson fields and ALP, are treated as classical fields where their field operators are replaced by their mean values.
The equations of motion of the mediators are
\begin{eqnarray}
    m_\sigma^2\sigma_0 &=& - g_\sigma\langle\bar{\psi}\psi\rangle - A\sigma_0^2 - B\sigma_0^3, \label{eq:eoms2} \\
    m_\omega^2 \omega_0 &=&  g_\omega\langle\bar{\psi}\gamma^0\psi\rangle, \\
    m_\rho^2 b_0 &=&  g_\rho \langle\bar{\psi}\gamma^0\tau_3\psi\rangle, \\
    m_a^2 a_0 &=& g_{a\chi\chi} \langle\bar{\chi} i \gamma^5 \chi\rangle + g_{aff} \langle\bar{\psi} i \gamma^5 \psi\rangle, \label{eq:eoma2}
\end{eqnarray}
and the equations of motion of nucleon and DM are
\begin{eqnarray}
    \left[ i\gamma^\mu\partial_\mu - g_\omega \gamma^0\omega_0 - g_\rho \gamma^0 \tau_3 b_0 - M^* - g_{aff} i \gamma^5  a_0 \right]\psi &=& 0, \label{eq:eomb2} \\
    \left(i\gamma^\mu\partial_\mu -m_\chi - g_{a\chi\chi} i \gamma^5 a_0 \right)\chi &=& 0, \label{eq:eomx2} 
\end{eqnarray}
where $M^*=(M + g_\sigma \sigma_0)$.
In order to estimate the energy of both nucleon and DM, we write the equations \ref{eq:eomb2} and \ref{eq:eomx2} in the matrix form and square both equations to avoid the non-diagonal component from $\gamma^5$.
The details of the calculation are provided in Appendix \ref{sect:app1}. We find that the effective energy of the nucleon and DM are
\begin{eqnarray}
    E_\psi &=& g_\omega \omega_0 + g_\rho \tau_3 b_0 \pm \sqrt{k^2 + (M^*)^2 +g_{aff}^2 a_0^2}, \label{eq:ef} \\
    E_\chi &=& \sqrt{q^2 + m_\chi^2 + g^2_{a\chi\chi} a_0^2}, \label{eq:ex}
\end{eqnarray}
where $k$ and $q$ are the momentum of the nucleon and DM, respectively. We can define the masses of the nucleon and DM as
\begin{eqnarray}
    \widetilde{m}^2 &=& (M^*)^2 + g_{aff}^2 a_0^2 = (M + g_\sigma \sigma_0)^2 + g_{aff}^2 a_0^2, \label{eq:m_t} \\
    \widetilde{m}_\chi^2 &=& m_\chi^2 + g^2_{a\chi\chi} a_0^2. \label{eq:m_x_t} 
\end{eqnarray}
The constants of the scalar meson, vector meson, isovector meson and ALP are given by
\begin{eqnarray}
    \sigma_0 &=& - \frac{g_\sigma}{m_\sigma^2} \frac{\gamma}{2\pi^2} \int_0^{k_f} dk \frac{k^2 M^*}{\sqrt{k^2 + \widetilde{m}^2}} - A\sigma_0^2 - B\sigma_0^3, \label{eq:phi0}\\
    \omega_0 &=& \frac{g_\omega}{m^2_\omega} \rho,  \label{eq:V0}\\
    b_0 &=& \frac{g_\rho}{m^2_\rho} \rho_3,  \label{eq:rho0}\\
    a_0 &=& \frac{g_{aff}^2}{m_a^2 } \frac{\gamma }{2\pi^2} a_0 \int_0^{k_f} dk \frac{k^2}{\sqrt{k^2 + \widetilde{m}^2} } \nonumber \label{eq:a0} \\
    & & + \frac{g_{a\chi\chi}^2}{m_a^2} \frac{\gamma_\chi }{2\pi^2} a_0 \int_0^{q_f} dq \frac{q^2}{\sqrt{q^2 + \widetilde{m}_\chi^2} },
\end{eqnarray}
where $k_f$ and $q_f$ are the Fermi momentum of nucleon and DM, and $\rho = \gamma  k_f^3/6\pi^2$ is the baryon density and $\rho_3 = (2y_p - 1)\rho$ where $y_p$ is the proton fraction. In general, the proton fraction can be determined by considering the conditions of beta equilibrium and charge neutrality. However, in this work, we focus on the case of pure neutron matter. The total energy density and pressure are given by 
\begin{eqnarray}
    \varepsilon &=& \frac{\gamma}{2\pi^2} \int_0^{k_f}dk\ k^2\sqrt{k^2 + \widetilde{m}^2} + \frac{\gamma_\chi}{2\pi^2} \int_0^{q_f}dq\ q^2\sqrt{q^2 + \widetilde{m}_\chi^2} + g_\omega \omega_0\rho +g_\rho b_0\rho_3 \nonumber \\
    & & + \frac{1}{2}m_\sigma^2\sigma_0^2 + \frac{1}{3}A\sigma_0^3 + \frac{1}{4}B\sigma_0^4 - \frac{1}{2}m_\omega^2 \omega_0^2 - \frac{1}{2}m_\rho^2 b_0^2 + \frac{1}{2}m_a^2 a_0^2, \label{eq:Etotal}\\ 
    P &=& \frac{\gamma}{6\pi^2} \int_0^{k_f} dk\frac{k^4}{\sqrt{k^2 + \widetilde{m}^2}} + \frac{\gamma_\chi}{6\pi^2} \int_0^{q_f} dq\frac{q^4}{\sqrt{q^2 + \widetilde{m}_\chi^2}} \nonumber \\
    & & - \frac{1}{2}m_\sigma^2\sigma_0^2 - \frac{1}{3}A\sigma_0^3 - \frac{1}{4}B\sigma_0^4 + \frac{1}{2}m_\omega^2 \omega_0^2 + \frac{1}{2}m_\rho^2 b_0^2 - \frac{1}{2}m_a^2 a_0^2. \label{eq:Ptotal}
\end{eqnarray}

We calculate the EoS of the system by numerically solving the equations \ref{eq:phi0}-\ref{eq:a0} simultaneously for the range of nucleon Fermi momenta, $k_f$, where $\gamma, \gamma_\chi=2$ are the spin degeneracy factors of the nucleon and DM. 
Then we substitute the values of mean fields into the equations \ref{eq:Etotal} and \ref{eq:Ptotal}. 

\begin{figure}[h]
    \centering
    \includegraphics[width=0.8\linewidth]{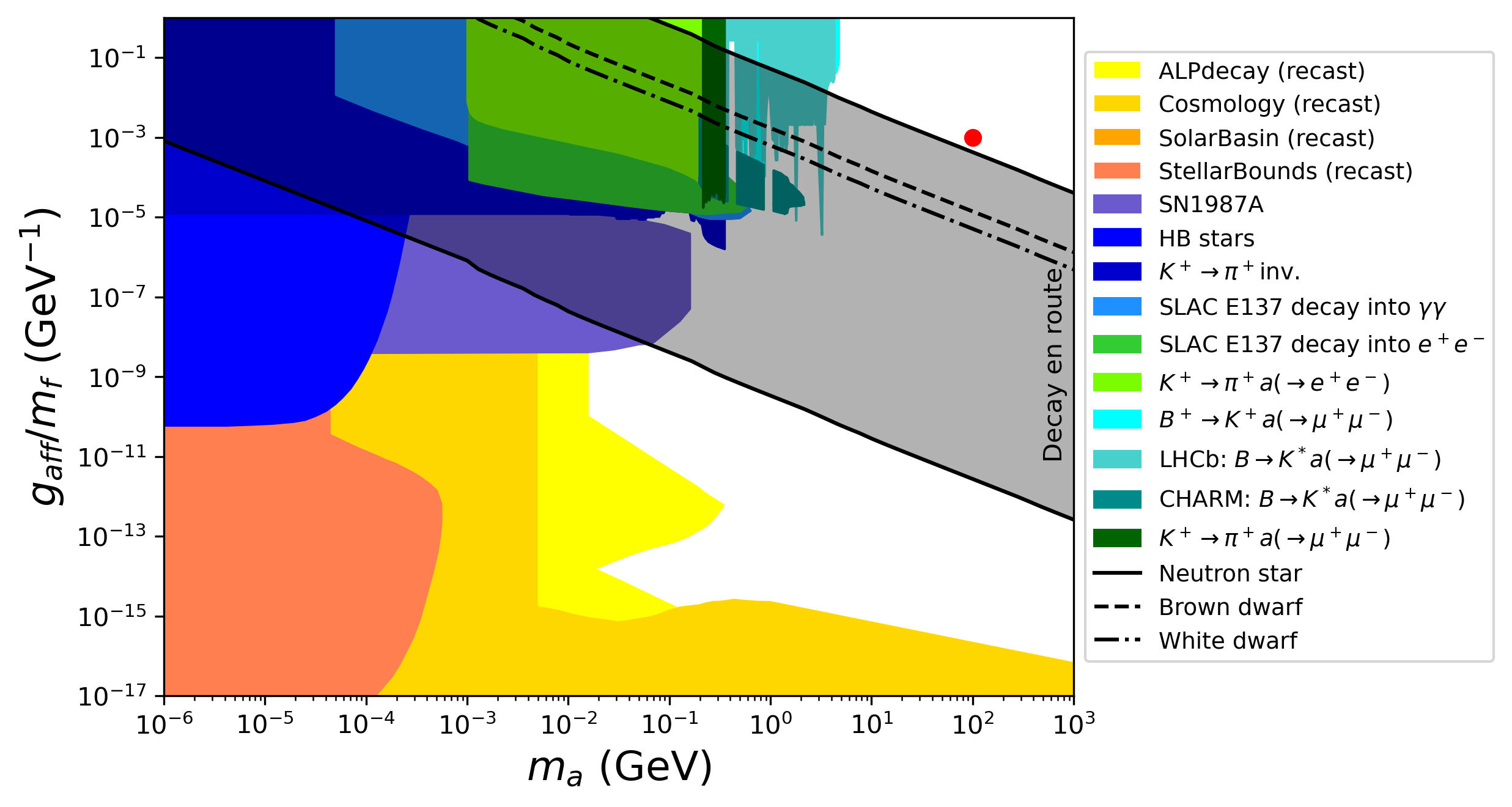}
    \caption{The plot shows results from our previous study \cite{Klangburam:2023vjv}. The upper boundary of the shaded gray region corresponds to the effective ALP coupling for the case where the ALP decays after escaping the surface of the compact object, while the lower boundary represents decay before reaching Earth. The red point indicates the benchmark parameters used in this work, with $m_a = 100$ GeV, $g_{aff} = 10^{-3}$, and $g_{a\chi\chi} = 10^{-3}$.
    }
    \label{fig:alp_decay}
\end{figure}

Using the parameters ($g_{aff},\ g_{a\chi\chi},\ m_a$) allowed by our previous study \cite{Klangburam:2023vjv} shown in Figure \ref{fig:alp_decay}, and relevant to this analysis, i.e., ALP cannot escape and cool down the neutron star, we found that the mean field of ALP becomes very small, as expected. For example, for $m_a = 10^2 - 10^3$ GeV and $g_{aff} = g_{a\chi\chi} = 10^{-3} - 1$ which allows the decay length of ALP to be shorter than the neutron star radius, the ALP mean field is between $a_0 = 10^{-10} - 10^{-8}$. As a result, the ALP will not contribute to any significant change in the EoS. Therefore, our study will focus only on the DM parameters. We investigate the impact of DM in two different cases, i.e., we fixed the values of $m_\chi$ as 200 and 500 GeV and varied the values of $q_f=0.02-0.05$ GeV or fixed $q_f$ as 0.02 and 0.04 GeV while varied $m_\chi=50-1000$ GeV. The EoS for different values of $m_\chi$ and $q_f$ are shown in the Figures \ref{fig:eos_mx} and \ref{fig:eos_qf}. The hadronic parameter was taken from the works of \cite{Lalazissis:1996rd,Das:2018frc}, where their numerical values are shown in Table \ref{tab:para1}.
\begin{table}[h]
    \centering
    \begin{tabular}{|cccccccc|}
    \hline
         $m_\sigma$ (MeV) & $g_\sigma$ & $m_\omega$ (MeV) & $g_\omega$ & $m_\rho$ & $g_\rho$ & A (fm$^{-1}$) & B \\ \hline
         508.194 & 10.217 & 782.501 & 12.868 & 763.000 & 4.474 & -10.431 & -28.885\\
    \hline
    \end{tabular}
    \caption{The NL3 parameter set for the hadronic part.}
    \label{tab:para1}
\end{table}

Although the QHD model considered in this work does not capture the full complexity of nuclear interactions, it provides a crucial foundation for analyzing the ALP-mediated DM scenario. A more sophisticated treatment of nuclear matter, along the lines of the approach \cite{Dutra:2014qga,Lopes:2021yga,Becerra:2024ond} will be studied in the future.

\section{TOV equations and tidal deformability}
\label{sect:3}

In this section, we describe the formalism that we use to study the properties of the neutron star. The metric for a static, spherically symmetric star is given by
\begin{eqnarray}
    ds^2 = -e^{2\nu(r)}dt^2 + e^{2\lambda(r)}dr^2 + r^2(d\theta^2 + \sin^2\theta d\phi^2), \label{eq:g}
\end{eqnarray}
where $\nu(r)$ and $\lambda(r)$ are the metric functions. After solving the Einstein field equation with the given metric, one can obtain the Tolman-Oppenheimer-Volkoff (TOV) equations \cite{Oppenheimer:1939ne,Tolman:1939jz} written as
\begin{eqnarray}
    \frac{dP}{dr} &=& - \frac{\left(4\pi r^3 P  + M\right)(\varepsilon + P)}{r( r - 2M)}, \label{eq:TOVp} \\
    \frac{dM}{dr} &=& 4 \pi r^2 \varepsilon. \label{eq:TOVm}
\end{eqnarray}
The coupled differential equations \ref{eq:TOVp} and \ref{eq:TOVm} can be numerically integrated from the center of the star to the surface. Given an EoS, the pressure at the center $(r=0)$ is $P_c = P(\varepsilon_c)$, while at the surface $(r=R)$, it reaches zero. By solving for all possible values of $\varepsilon_c$ from the EoS, we determine the stellar mass, $M = M_\star$, and radius, $R = R_\star$. The mass-radius relation then reveals the maximum mass, beyond which a neutron star becomes unstable and collapses into a black hole.

The tidal deformability quantifies the quadrupole deformation of a compact object in a binary system due to the tidal effect of its companion star \cite{Thorne:1997kt,Hinderer:2007mb,Flanagan:2007ix,Hinderer:2009ca,Binnington:2009bb,Damour:2009vw}. It can be interpreted as the $l=2$ perturbation of a spherically symmetric star, with the perturbation metric given by \cite{Regge:1957td,Hinderer:2007mb}
\begin{eqnarray}
    h_{\mu\nu} = \text{diag}\left[e^{2\nu(r)}H_0(r), e^{2\lambda(r)}H_2(r), r^2 K(r), r^2\sin^2\theta K(r)  \right] Y_{2m}(\theta,\phi).  \label{eq:h}
\end{eqnarray}

Details of the tidal deformability calculation for neutron stars can be found in \cite{Hinderer:2007mb}. The relation between the induced quadrupole moment tensor and the tidal field tensor is given by \cite{Flanagan:2007ix,Hinderer:2009ca}, $Q_{ij}= - \lambda \mathcal{E}_{ij}$, where $\lambda$ is related to the tidal Love number $(l=2)$, $k_2$, as

\begin{eqnarray}
    \lambda = \frac{2}{3}k_2 R^5, \label{eq:lambda}
\end{eqnarray}
with $R$ being the radius of the neutron star. The parameter $\lambda$ quantifies the degree of deformation due to the external tidal field and increases with the radius of the neutron star. The dimensionless tidal deformability is defined as

\begin{eqnarray}
    \Lambda = \frac{\lambda}{M^5} = \frac{2 k_2}{3 C^5},  \label{eq:lambda_dimless}
\end{eqnarray}
where $C\equiv M/R$ denotes the compactness of the star. The tidal Love number is given by \cite{Hinderer:2009ca,Damour:2009vw,Postnikov:2010yn}

\begin{eqnarray}
    k_2 &=& \frac{8C^5}{5}(1-2C^2)[2 + 2C(y-1) - y]\times \nonumber \\
    & & \Big\{ 2C(6 - 3y + 3C(5y - 8)) + 4C^3[13 - 11y + C(3y-2) +2C^2(1+y)] \nonumber \\
    & & + 3(1-2C)^2[2 - y + 2C(y - 1)]\log{(1-2C)} \Big\}^{-1}, \label{eq:k2}
\end{eqnarray}
 where $y=r H'(r)/H(r)$ is obtain by solving the following differential equation \cite{Hinderer:2009ca,Damour:2009vw,Postnikov:2010yn}
\begin{eqnarray}
    ry'(r) + y(r)^2 + y(r)F(r) + r^2Q(r) = 0, \label{eq:ODEtidal}
\end{eqnarray}
evaluated at $r=R$. The functions $F(r)$ and $Q(r)$ are defined as
\begin{eqnarray}
    F(r) &=& \left[ 1 + 4\pi r^2(P-\varepsilon) \right]\left( 1 - \frac{2M}{r} \right)^{-1},  \\
    Q(r) &=& 4\pi \left[5\varepsilon + 9P + \frac{\varepsilon + P}{dp/d\varepsilon}\right] \left( 1 - \frac{2M}{r} \right)^{-1} - \frac{6}{r^2}\left( 1 - \frac{2M}{r} \right)^{-1} \nonumber \\
    & & -\frac{4M^2}{r^4} \left(1 + \frac{4\pi r^3 P}{M}\right)^2 \left( 1 - \frac{2M}{r} \right)^{-2}.
\end{eqnarray}

For a given EoS, we numerically solve equations \ref{eq:TOVp} and \ref{eq:TOVm} together with equation \ref{eq:ODEtidal} to determine the stellar mass $M$, radius $R$, and the dimensionless tidal deformability $\Lambda$. The mass-radius diagrams for the different EoS are shown in Figures \ref{fig:MR_mx} and \ref{fig:MR_qf}, and the tidal deformability-mass diagrams are shown in Figures \ref{fig:Lambda_mx} and \ref{fig:Lambda_qf}.

\section{Results}
\label{sect:4}

In this work, we investigate the impact of dark matter (DM) on neutron stars using the ALP-mediated DM model. We consider two scenarios: one where the DM mass is fixed and another where the DM Fermi momentum is fixed. In the first case, we set the DM mass to 200 GeV and 500 GeV while varying the DM Fermi momentum, $q_f$, from 
0.02 GeV to 0.05 GeV. In the second case, we fix the DM Fermi momentum at $q_f = 0.02$ GeV and 0.04 GeV while varying the DM mass from 
50 GeV to 1000 GeV. In both cases, the ALP mass is set to 
$m_a = 100$ GeV, with couplings $g_{aff} = 10^{-3}$ and $g_{a\chi\chi} = 10^{-3}$. 
In this work, we use a set of parameters for the ALP couplings taken from reference~\cite{Klangburam:2023vjv} with the assumption that $g_{aff} \ll g_{a\gamma\gamma}$.
This parameter set corresponds to a region where the ALP has the decay length shorter than the radius of the neutron star and is relevant for the physical scenario considered here.

Using these parameter sets, we compute the equation of state (EoS) via equations \ref{eq:Etotal} and \ref{eq:Ptotal}. Once the EoS is obtained, we numerically solve equations \ref{eq:TOVp}, \ref{eq:TOVm}, and \ref{eq:ODEtidal} simultaneously to derive the mass-radius and tidal deformability-mass relations. The results are presented alongside the purely hadronic neutron star case, denoted as ``no DM'', represented by the black line. 

\begin{figure}[h]
\centering
\begin{subfigure}{0.49\textwidth}
    \includegraphics[width=\linewidth]{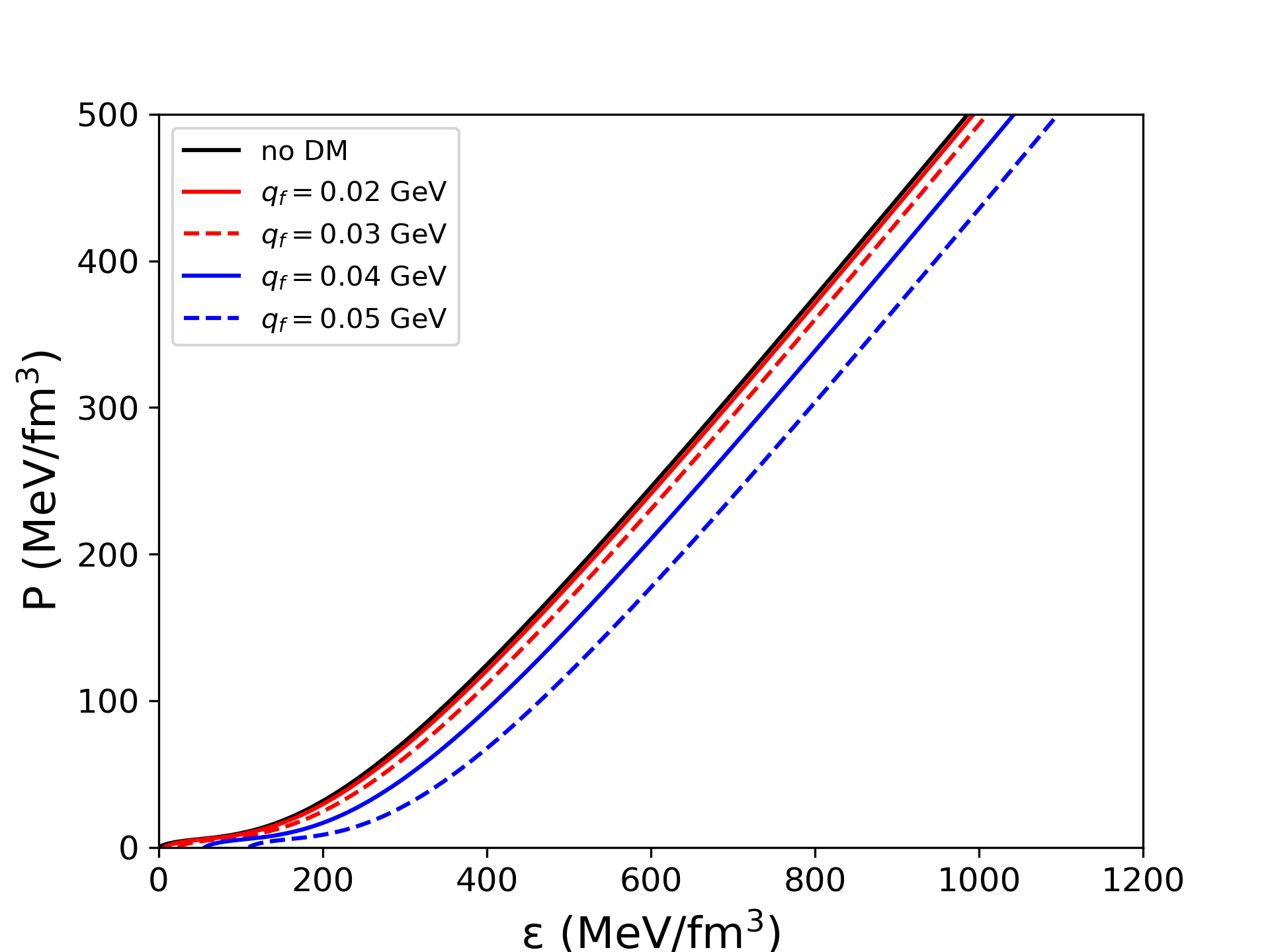}
    \caption{\footnotesize $m_\chi=200$ GeV}
    \label{fig:eos_mx_200}
\end{subfigure}
\begin{subfigure}{0.49\textwidth}
    \includegraphics[width=\linewidth]{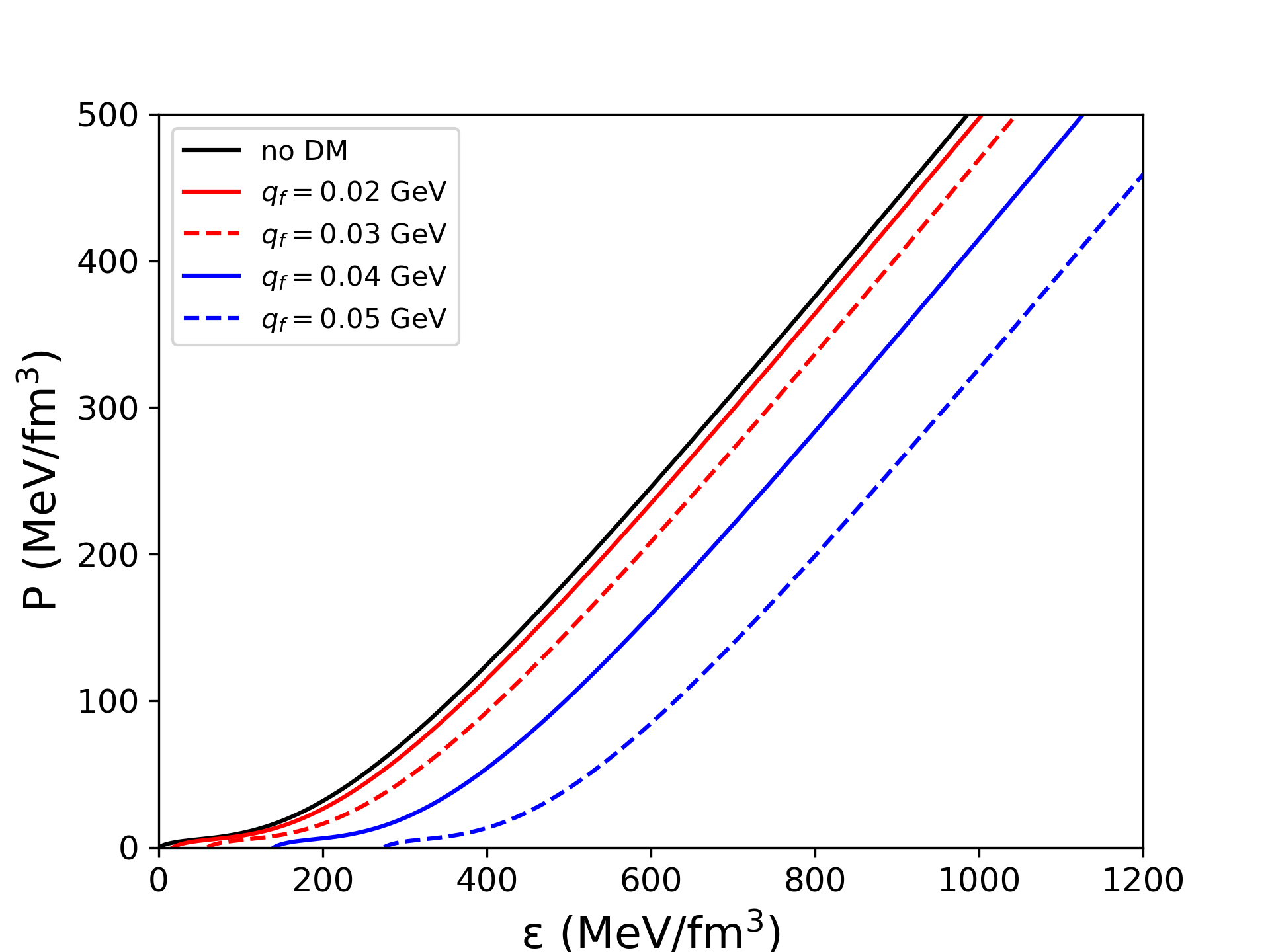}
    \caption{\footnotesize $m_\chi=500$ GeV}
    \label{fig:eos_mx_500}
\end{subfigure}
        
\caption{The EoSs of the system where we fixed $m_\chi=$ 200 and 500 GeV. The other parameters of DM model are $m_a = 100$ GeV, $g_{aff} = 10^{-3}$ and $g_{a\chi\chi} = 10^{-3}$.} 
\label{fig:eos_mx}
\end{figure}

\begin{figure}[h]
\centering
\begin{subfigure}{0.49\textwidth}
    \includegraphics[width=\linewidth]{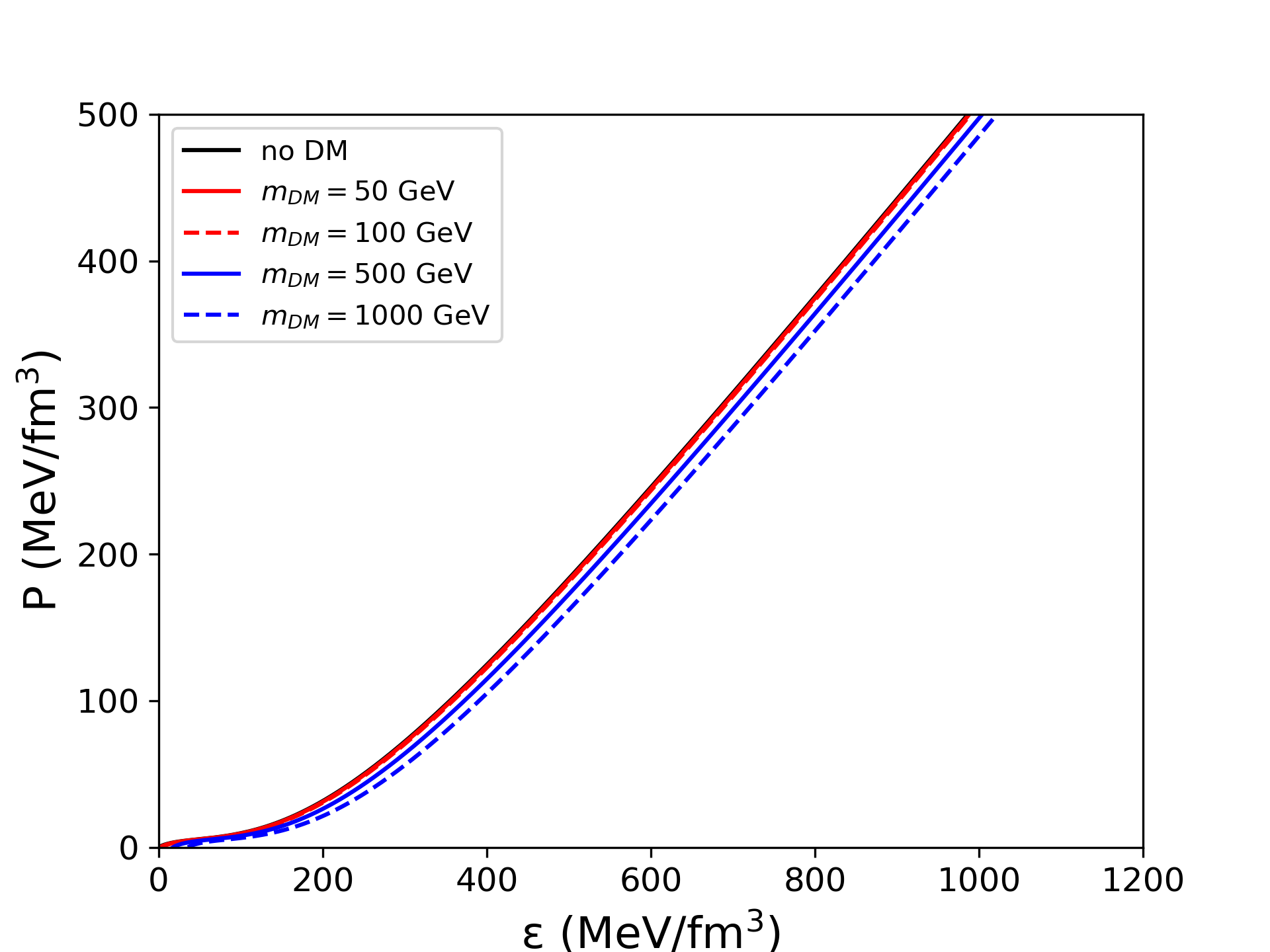}
    \caption{\footnotesize $q_f = 0.02$ GeV}
    \label{fig:eos_qf_0.02}
\end{subfigure}
\begin{subfigure}{0.49\textwidth}
    \includegraphics[width=\linewidth]{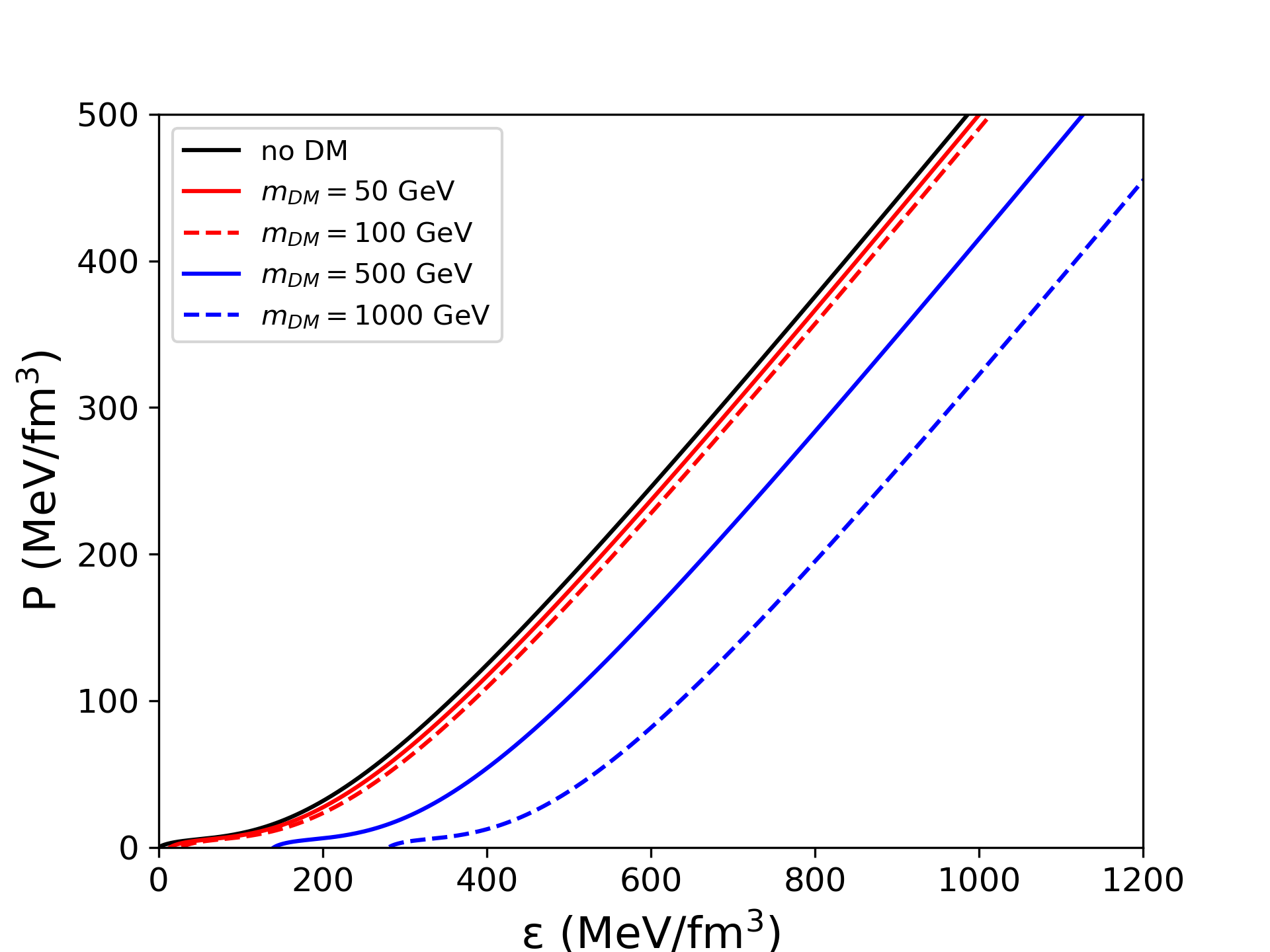}
    \caption{\footnotesize $q_f = 0.04 $ GeV}
    \label{fig:eos_qf_0.04}
\end{subfigure}
        
\caption{The EoSs of the system where we fixed $q_f =$ 0.02 and 0.04 GeV. The other parameters of DM model are $m_a = 100$ GeV, $g_{aff} = 10^{-3}$ and $g_{a\chi\chi} = 10^{-3}$.} 
\label{fig:eos_qf}
\end{figure}

Figure \ref{fig:eos_mx} presents the EoSs for different values of $q_f$ with a fixed DM mass: $m_\chi = 200$ GeV (Figure \ref{fig:eos_mx_200}) and $m_\chi = 500$ GeV (Figure \ref{fig:eos_mx_500}). 
The results indicate that increasing the DM Fermi momentum softens the EoS, implying a reduction in pressure support and shifting the EoS to higher energy densities. This leads to a lower maximum mass for the neutron star. Moreover, the impact of $q_f$ is more pronounced at higher DM masses. In Figure \ref{fig:eos_qf}, we show the EoSs for fixed $q_f$ values of 0.02 GeV (Figure \ref{fig:eos_qf_0.02}) and 0.04 GeV (Figure \ref{fig:eos_qf_0.04}), with varying DM masses. For small $q_f$, increasing the DM mass has little effect As in Figure.~2a, the results for $q_f = 0.02$ GeV and 0.03 GeV are indistinguishable from the case of the ``no DM". However, for larger $q_f$, only high DM masses (500 GeV and 1000 GeV) significantly soften the EoS. Note that this analysis considers only the neutron star core and neglects crust contributions. 

\begin{figure}[h]
\centering
\begin{subfigure}{0.49\textwidth}
    \includegraphics[width=\linewidth]{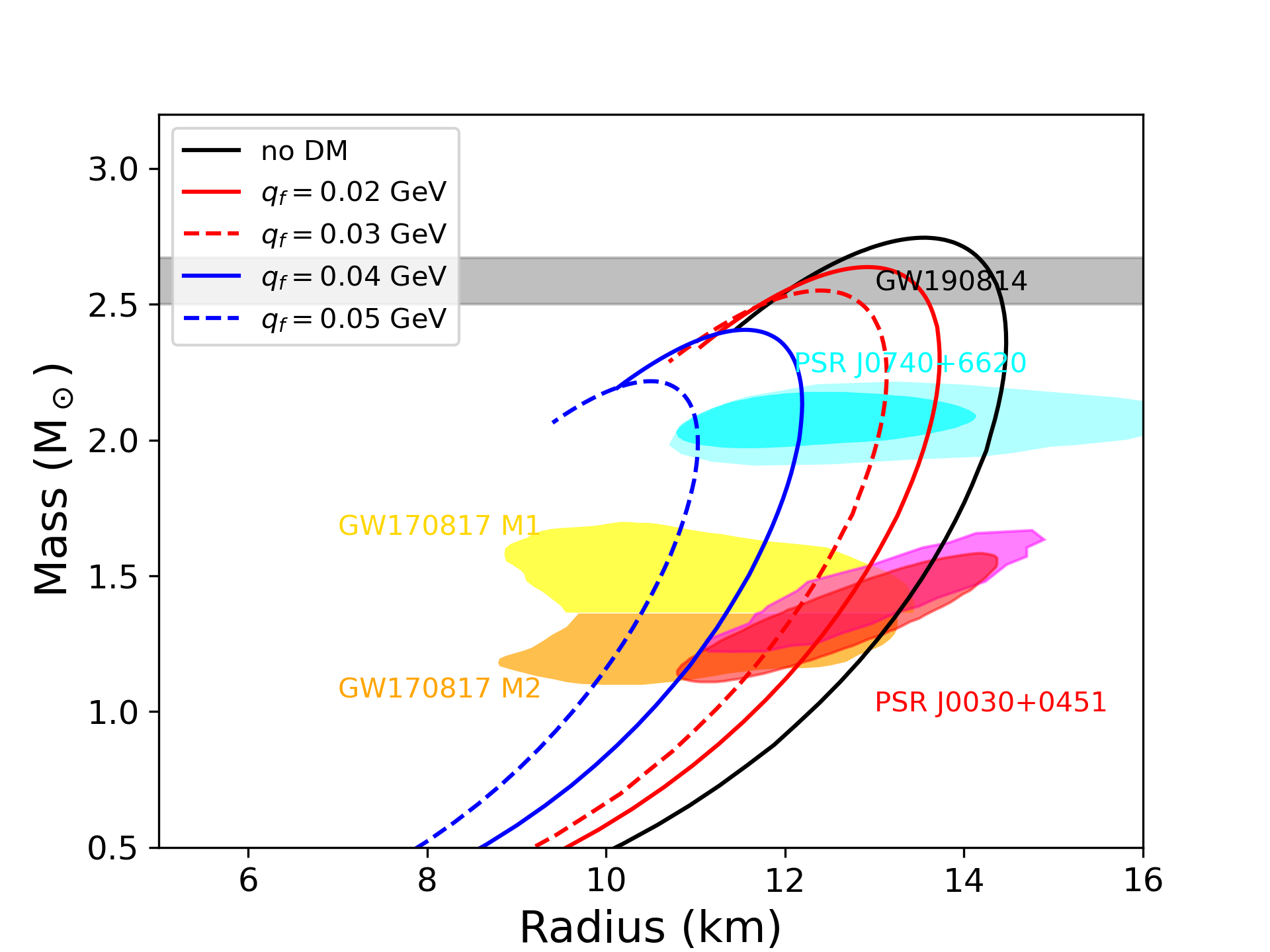}
    \caption{\footnotesize $m_\chi=200$ GeV}
    \label{fig:MR_mx_200}
\end{subfigure}
\begin{subfigure}{0.49\textwidth}
    \includegraphics[width=\linewidth]{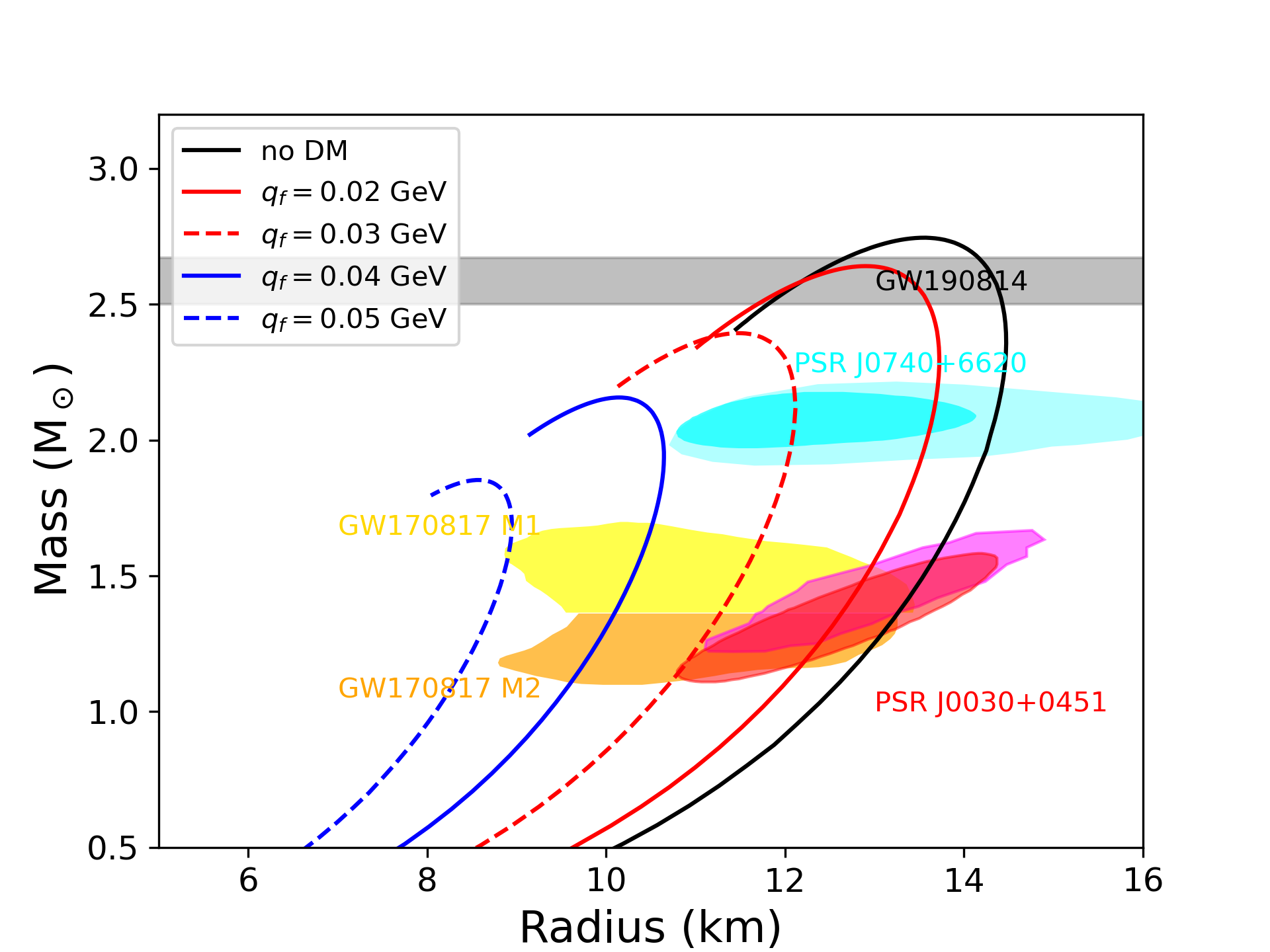}
    \caption{\footnotesize $m_\chi=500$ GeV}
    \label{fig:MR_mx_500}
\end{subfigure}
        
\caption{The mass-radius diagrams where we fixed $m_\chi=$ 200 and 500 GeV. The other parameters of DM model are $m_a = 100$ GeV, $g_{aff} = 10^{-3}$ and $g_{a\chi\chi} = 10^{-3}$.} 
\label{fig:MR_mx}
\end{figure}

\begin{figure}[h]
\centering
\begin{subfigure}{0.49\textwidth}
    \includegraphics[width=\linewidth]{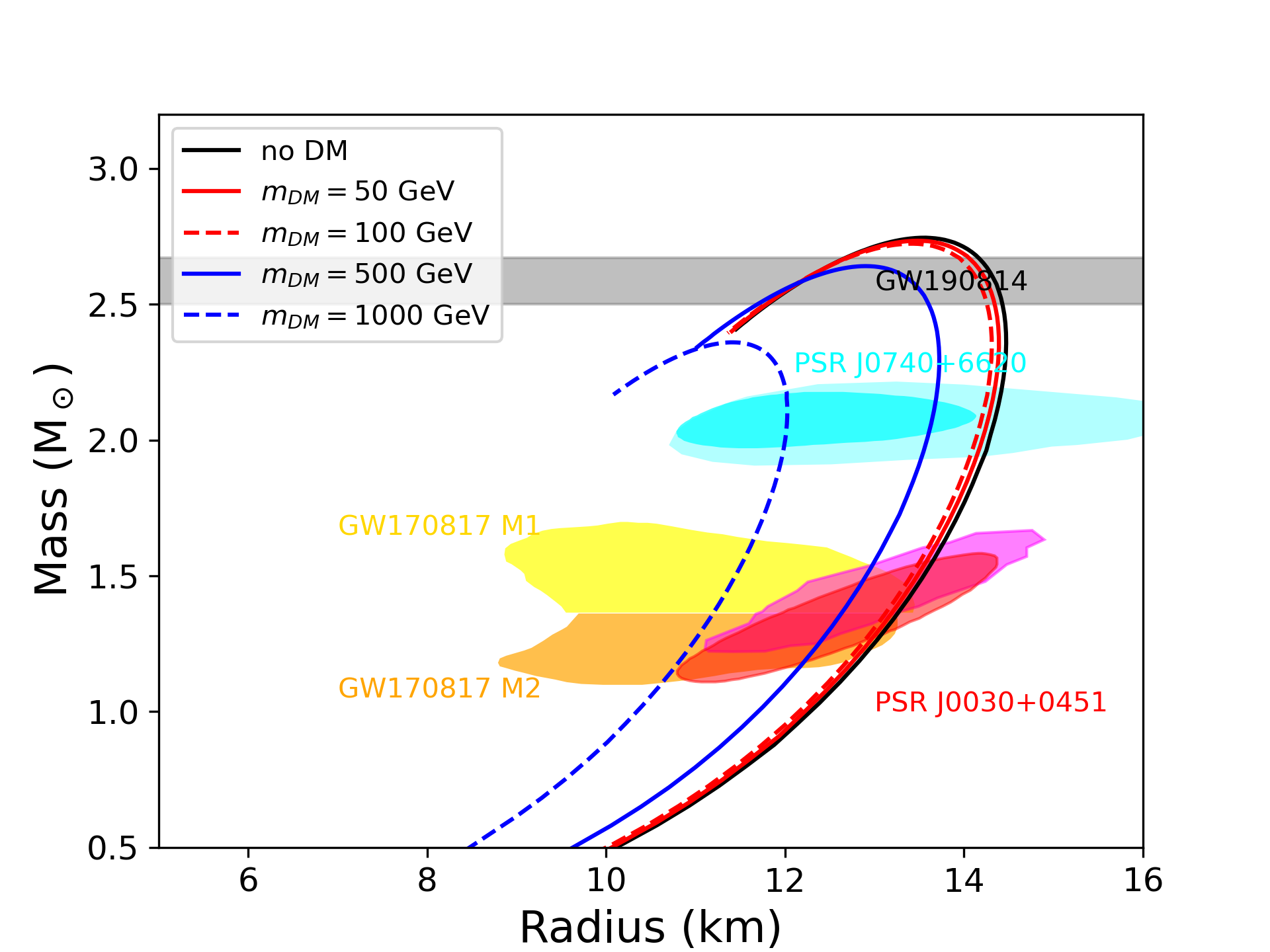}
    \caption{\footnotesize $q_f = 0.02$ GeV}
    \label{fig:MR_qf_0.02}
\end{subfigure}
\begin{subfigure}{0.49\textwidth}
    \includegraphics[width=\linewidth]{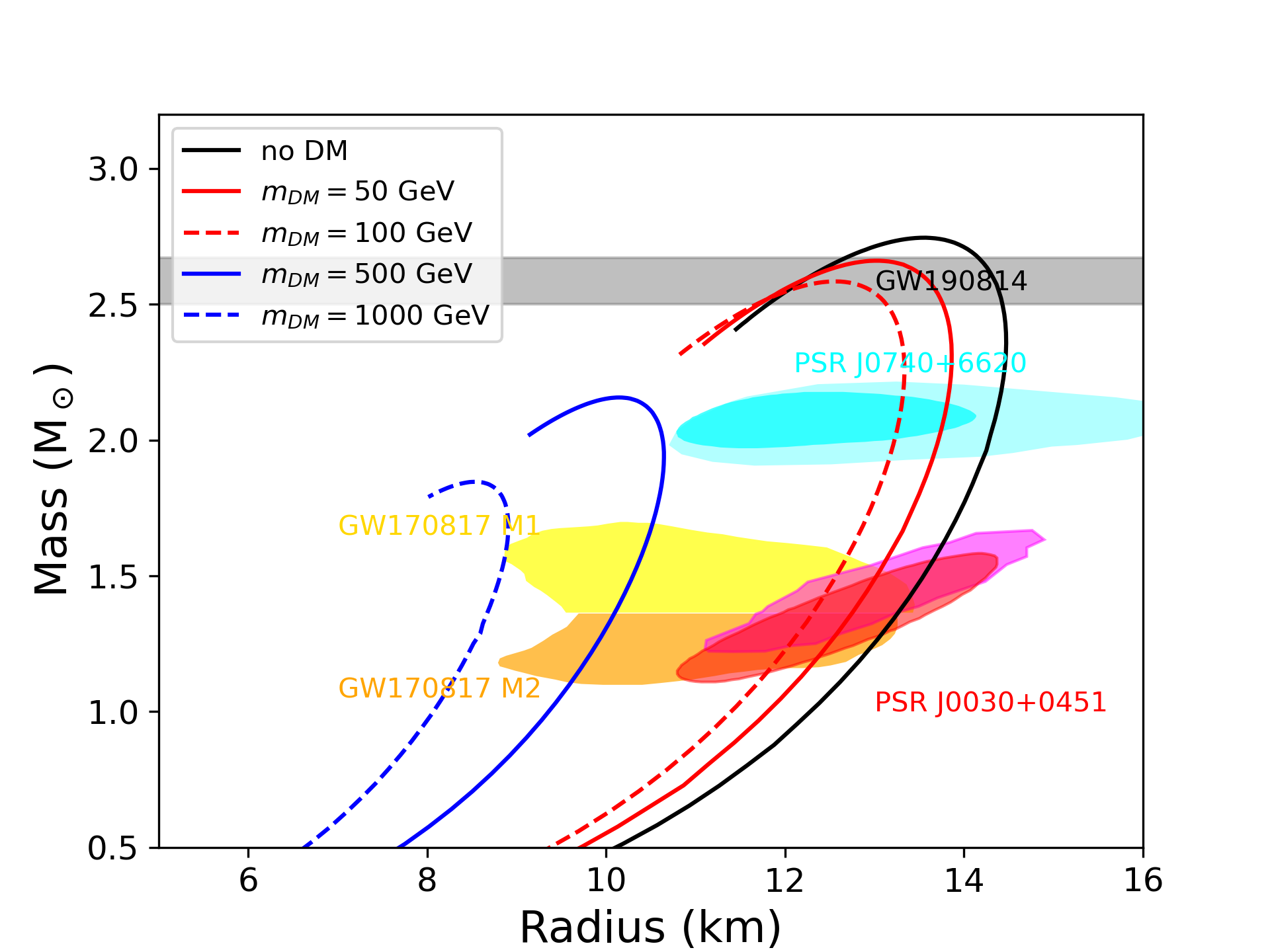}
    \caption{\footnotesize $q_f = 0.04$ GeV}
    \label{fig:MR_qf_0.04}
\end{subfigure}

\caption{The mass-radius diagrams where we fixed $q_f =$ 0.02 and 0.04 GeV. The other parameters of DM model are $m_a = 100$ GeV, $g_{aff} = 10^{-3}$ and $g_{a\chi\chi} = 10^{-3}$.} 
\label{fig:MR_qf}
\end{figure}

Figures \ref{fig:MR_mx} and \ref{fig:MR_qf} show the mass-radius relation for the first and second cases, respectively. We display the constraints from the analysis of the event GW170817 \cite{LIGOScientific:2018cki} in the yellow (GW170817 M1) and orange (GW170817 M2) regions. The constraints from the NICER experiment of PSR J0740+6620 \cite{Miller:2021qha,Riley:2021pdl} and PSR J0030+0451 \cite{Riley:2019yda,Miller:2019cac} are depicted in cyan and red regions, respectively. The constraint on the mass of the secondary component of GW190814 \cite{LIGOScientific:2020zkf} is displayed by the gray band. Figures \ref{fig:MR_mx} and \ref{fig:MR_qf} illustrate that increasing either the DM Fermi momentum or DM mass reduces both the maximum mass $(M_{max})$ and the radius of the neutron star. 
For a smaller fixed DM mass (Figure \ref{fig:MR_mx_200}), the results for $q_f=0.02$ GeV and $0.03$ GeV appear to satisfy all observational constraints. The results for $q_f=0.04$ GeV fails to meet the constraints from GW170817, while $q_f=0.05$ GeV fails to satisfy the constraints from GW170817 and PSR J0030+0451. For a higher DM mass (Figure \ref{fig:MR_mx_500}), the impact of DM Fermi momentum becomes more evident comparing with the results from a smaller value of DM mass. 
The result for $q_f = 0.02$ GeV is consistent with all the observational constraints. For $q_f=0.03$ GeV, the result fails to meet the constraints from GW190814. The case with $q_f=0.04$ GeV satisfies only the constraint from GW170817, while $q_f=0.05$ only slightly satisfies the constraints from GW170817. In Figure \ref{fig:MR_qf_0.02} where we fixed $q_f=0.02$ GeV, the impact of DM mass slightly reduces the maximum mass and the radius of the neutron star, with a more noticeable effect at higher DM mass (500 GeV and 1000 GeV). However, all results for $q_f=0.02$ GeV remain consistent with observational constraints, while only the case with $q_f=0.05$ GeV fails to satisfy the constraints from GW190814. For a higher fixed value of DM Fermi momentum (Figure \ref{fig:MR_qf_0.04}), the results for 
$m_\chi = 50$ GeV and $100$ GeV satisfy all observational constraints. However, the result for $m_\chi = 500$ GeV fails to meet the constraints from GW190814 and PSR J0030+0451, while the case with $m_\chi = 1000$ GeV is only slightly consistent with GW170817. 

\begin{figure}[h]
    \centering
    \includegraphics[width=0.6\linewidth]{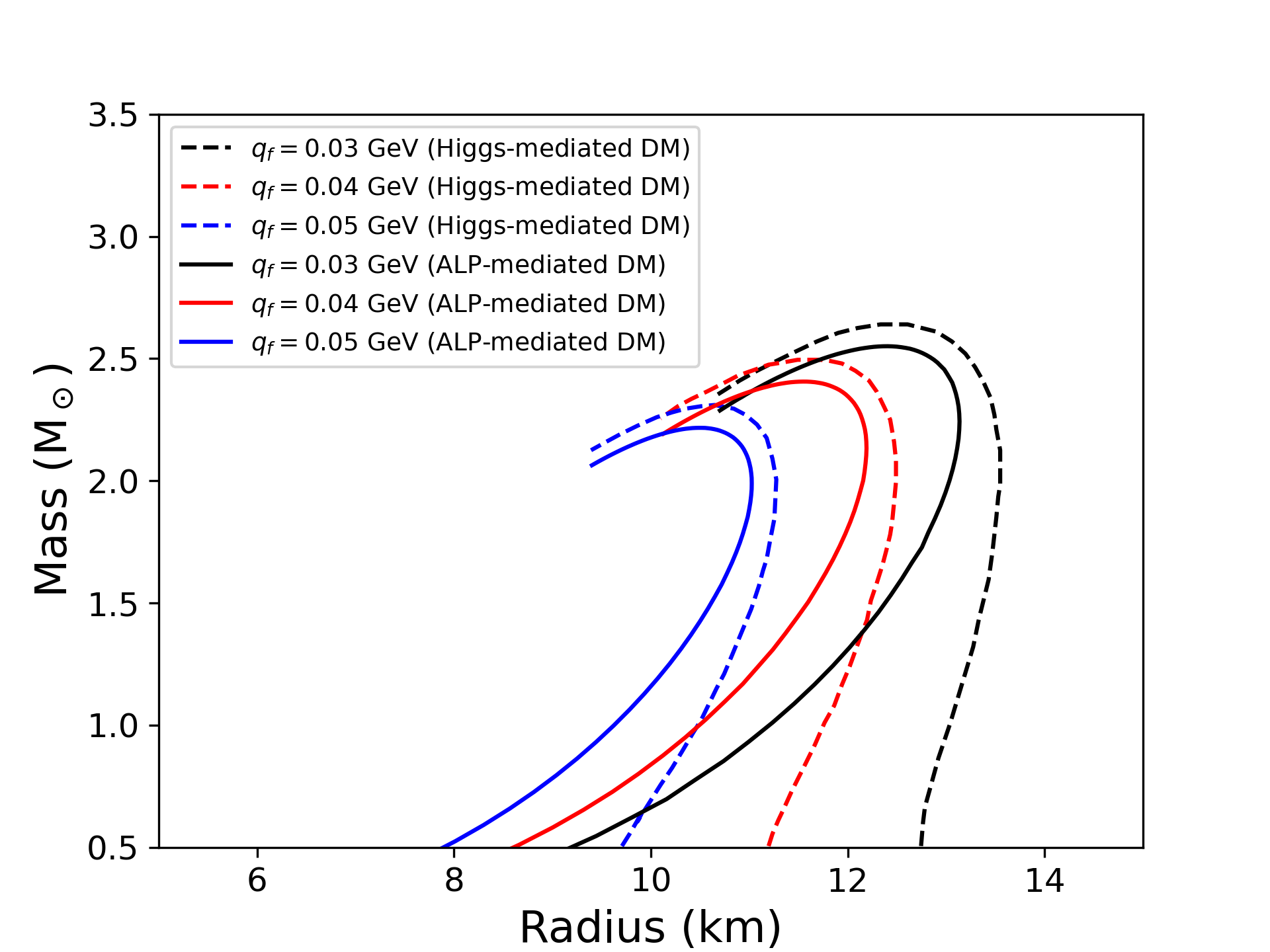}
    \caption{The plot shows the comparison of the mass-radius relations between Higgs and ALP mediator models with $m_\chi =200$ GeV. The dashed lines correspond to the Higgs mediator case, where their values are taken from \cite{Das:2018frc}, while the solid lines represent the results for the ALP mediator model.
    }
    \label{fig:alp_NL3}
\end{figure}

Finally, we also compare our mass-radius result with a more traditional Higgs-mediated model studied in \cite{Das:2018frc}. It turns out that the maximum mass and maximum radius of our model are slightly lower than those of Higgs-mediated model as shown in Figure \ref{fig:alp_NL3}. 

\begin{figure}[h]
\centering
\begin{subfigure}{0.49\textwidth}
    \includegraphics[width=\linewidth]{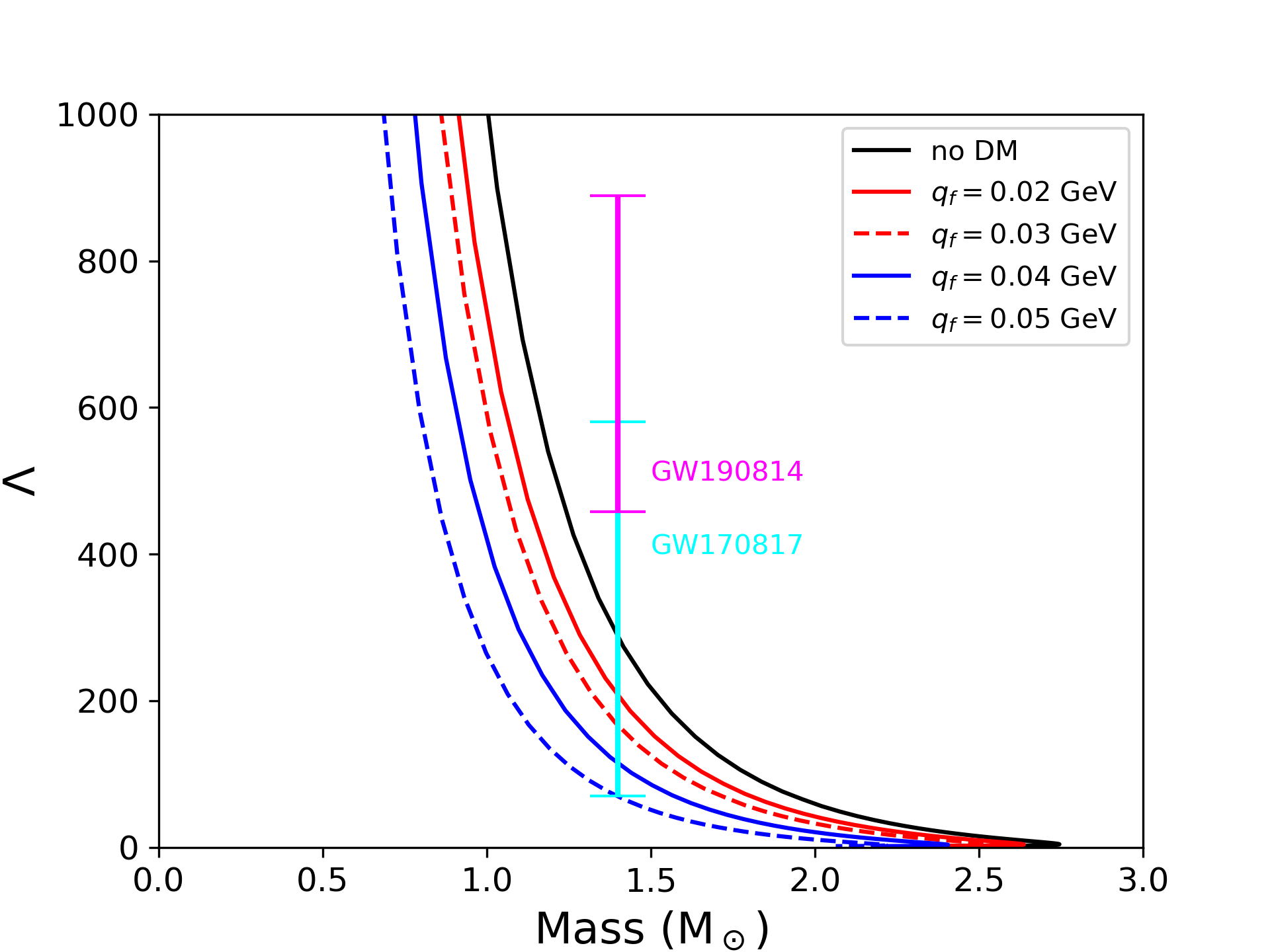}
    \caption{\footnotesize $m_\chi=200$ GeV}
    \label{fig:Lambda_mx_200}
\end{subfigure}
\begin{subfigure}{0.49\textwidth}
    \includegraphics[width=\linewidth]{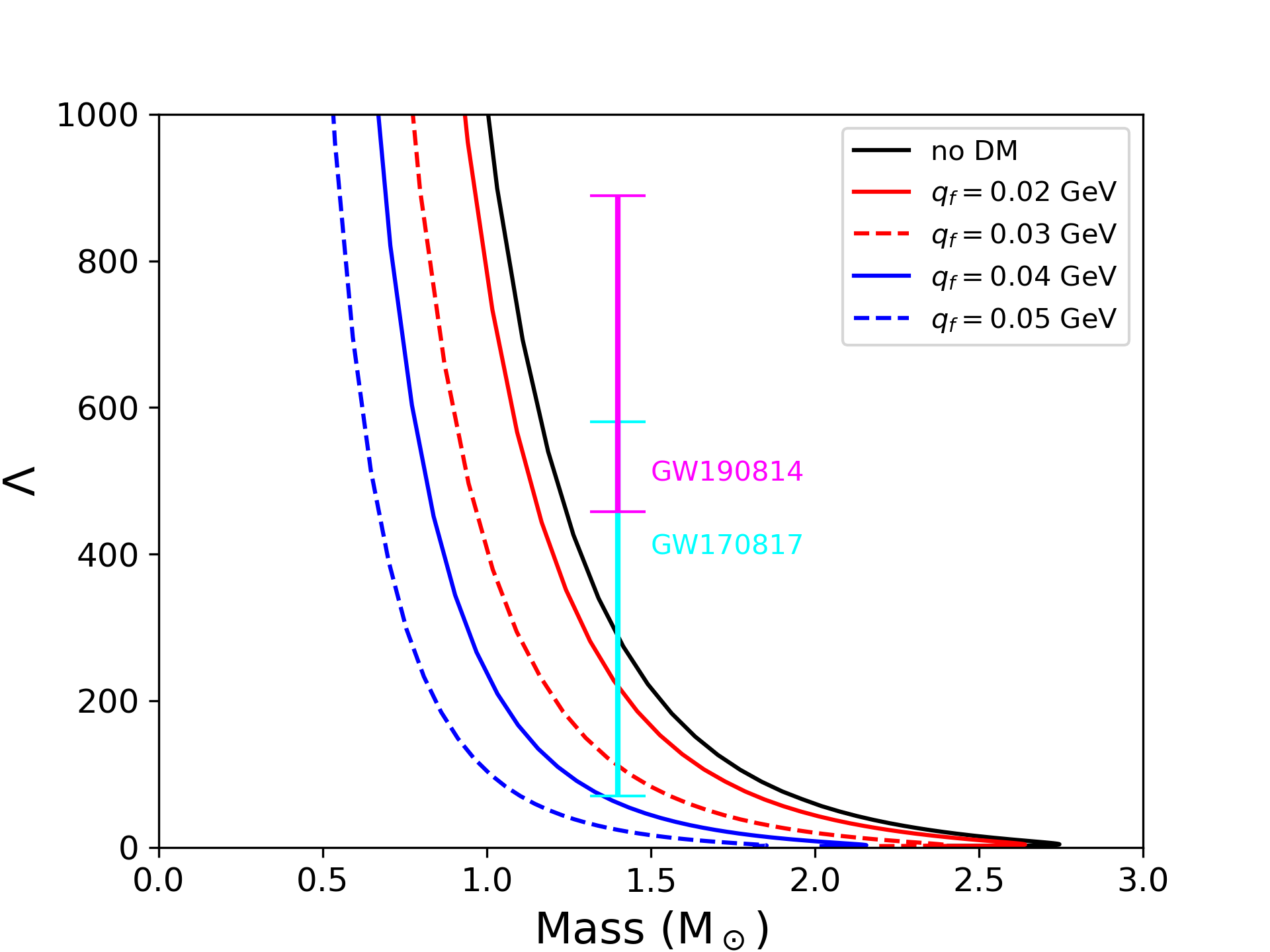}
    \caption{\footnotesize $m_\chi=500$ GeV}
    \label{fig:Lambda_mx_500}
\end{subfigure}
        
\caption{The tidal deformability-mass diagrams where we fixed $m_\chi=$ 200 and 500 GeV. The other parameters of DM model are $m_a = 100$ GeV, $g_{aff} = 10^{-3}$ and $g_{a\chi\chi} = 10^{-3}$. } 
\label{fig:Lambda_mx}
\end{figure}

\begin{figure}[h]
\centering
\begin{subfigure}{0.49\textwidth}
    \includegraphics[width=\linewidth]{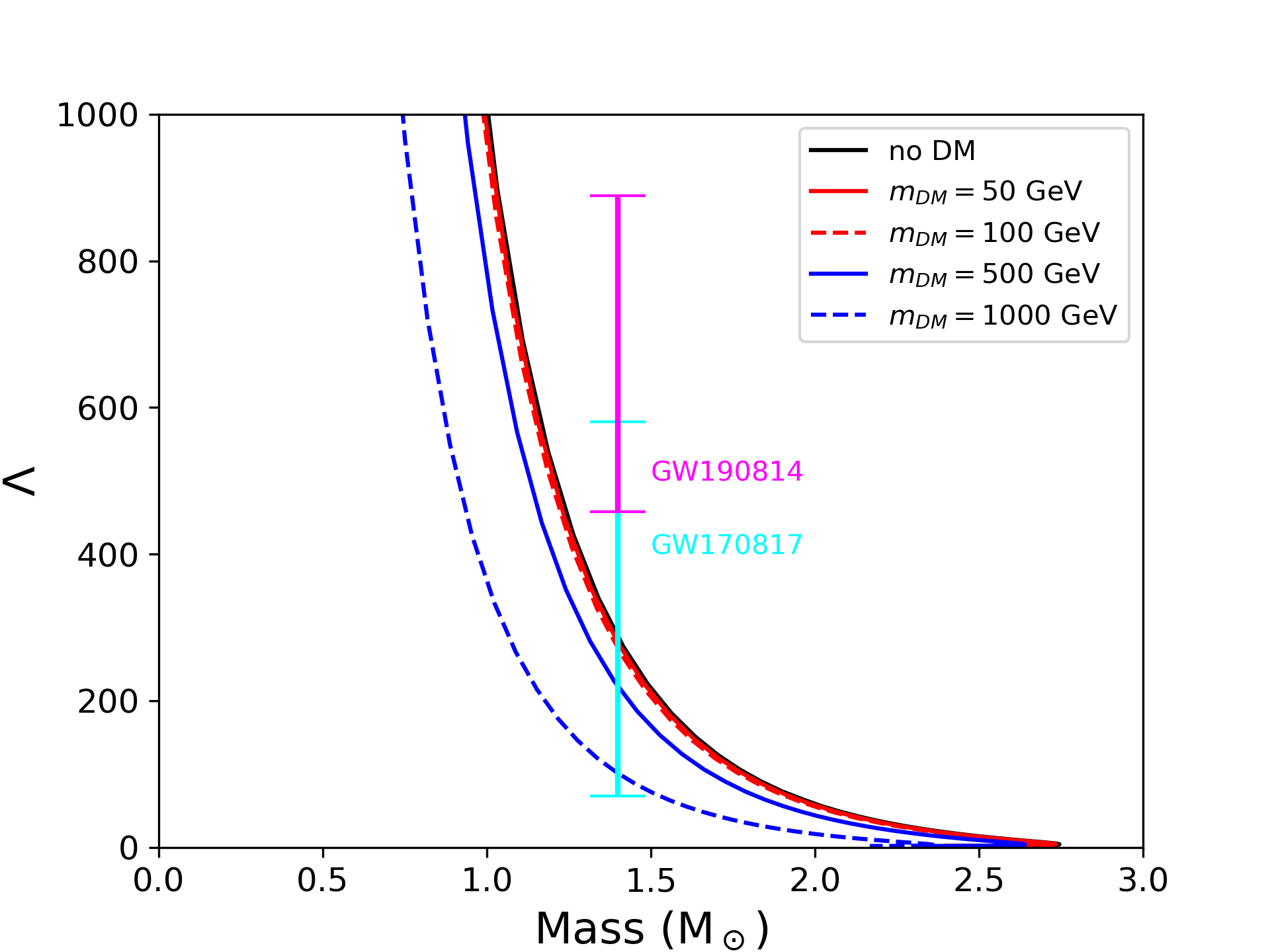}
    \caption{\footnotesize $q_f = 0.02$ GeV}
    \label{fig:Lambda_qf_0.02}
\end{subfigure}
\begin{subfigure}{0.49\textwidth}
    \includegraphics[width=\linewidth]{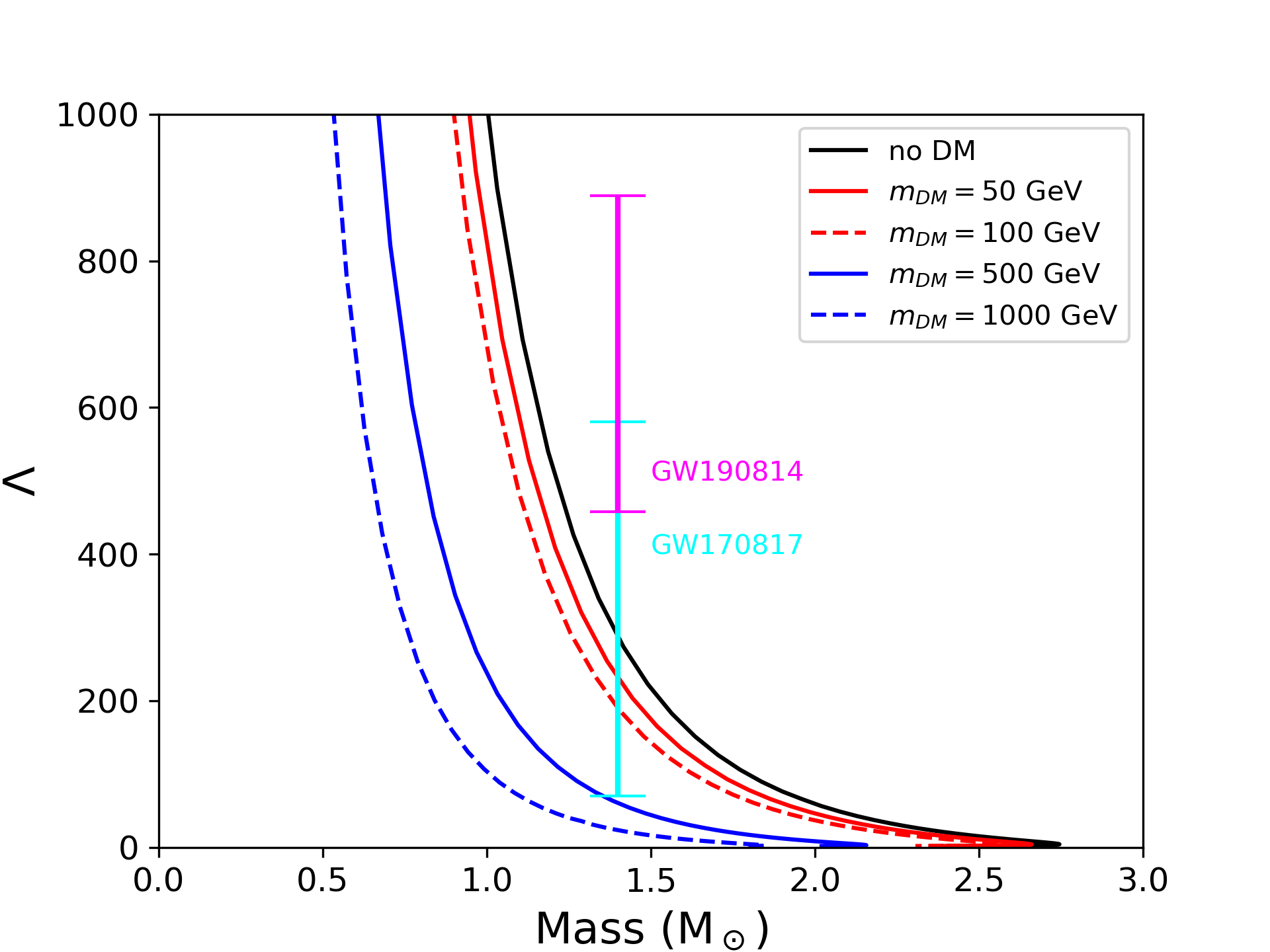}
    \caption{\footnotesize $q_f = 0.04$ GeV}
    \label{fig:Lambda_qf_0.04}
\end{subfigure}
        
\caption{The tidal deformability-mass diagrams where we fixed $q_f =$ 0.02 and 0.04 GeV. The other parameters of DM model are $m_a = 100$ GeV, $g_{aff} = 10^{-3}$ and $g_{a\chi\chi} = 10^{-3}$.} 
\label{fig:Lambda_qf}
\end{figure}

The results for the tidal deformability-mass relation are presented in Figures \ref{fig:Lambda_mx} and \ref{fig:Lambda_qf}. We also display the constraints on the tidal deformability of a neutron star with $1.4$ M$_\odot$ from GW170817 \cite{LIGOScientific:2018cki} $(\Lambda_{1.4} = 190^{+390}_{-120})$ and GW190814 \cite{LIGOScientific:2020zkf} $(\Lambda_{1.4} = 616^{+273}_{-158})$, represented by cyan and magenta lines, respectively. For a smaller fixed DM mass (Figure \ref{fig:Lambda_mx_200}), the results satisfy the GW170817 constraint for 
$q_f = 0.02-0.04$ GeV. 
At a higher fixed DM mass (Figure \ref{fig:Lambda_mx_500}), the results for 
$q_f = 0.02$ GeV and $0.03$ GeV remain consistent with GW170817, while $q_f = 0.04$ GeV and $0.05$ GeV fail to meet tidal deformability constraints.
In Figure \ref{fig:Lambda_qf_0.02}, where we fix $q_f = 0.02$ GeV, 
the results for $m_\chi = 50-1000$ GeV remain consistent with GW170817 where the result of the smallest value of DM mass ($m_\chi=50$ GeV) is very similar to the case without DM. 
For a higher fixed DM Fermi momentum (Figure \ref{fig:Lambda_qf_0.04}), the case of $m_\chi = 50$ GeV and $100$ GeV satisfy the tidal deformability constraint of GW170817. 
However, it should be noted that none of the results across the considered parameter space satisfy the constraint from GW190814.

\section{Discussion and Conclusions}
\label{sect:5}

In this work, we have studied the effects of the ALP-mediated dark matter (DM) model on the equation of state (EoS) of neutron stars. In this model, the interaction between nucleons and DM is mediated by an axion-like particle (ALP). We applied the relativistic mean-field approximation to the QHD-ALP-DM system and computed the EoS of neutron stars. Motivated by our previous study \cite{Klangburam:2023vjv}, we found that typical ALP parameter values do not lead to significant changes in the EoS. While the general setup bears similarity to earlier works on DM-admixed neutron stars, our approach offers complementary study of our previous analysis with a different phenomenology, i.e, gamma-ray and neutrino observation. Therefore, we only focused on investigating the impact of DM parameters, namely the DM mass ($m_\chi$) and DM Fermi momentum ($q_f$), on the EoS, mass-radius relation, and tidal deformability of neutron stars. 

Our analysis shows that increasing DM content shifts the energy density to higher values. This is because the presence of DM inside neutron stars modifies their internal structure by introducing additional degrees of freedom that soften the overall pressure support. Since DM does not participate in standard nuclear interactions, its inclusion dilutes the hadronic pressure, lowering both the maximum mass and radius of the neutron star. As a result, neutron stars with significant DM content become more compact and less deformable under tidal forces.  

By comparing our model with observational constraints from gravitational wave events and pulsar measurements, we find that the allowed parameter space for this model is limited to $q_f \leq 0.05$ GeV and $m_{\chi} \leq 1000$ GeV. TeV-scale DM highlights the importance of future multi-messenger observations in constraining the role of DM in neutron stars. In particular, next-generation gamma-ray observatories, such as the Cherenkov Telescope Array (CTA) \cite{CTAConsortium:2017dvg,CTA:2020qlo,Pinchbeck:2024xpc}, could provide valuable insights by probing potential high-energy signatures from neutron stars containing ALP-mediated DM.

In future work, we aim to extend our analysis by incorporating hyperons into the nuclear matter sector to achieve a more comprehensive description of dense matter interactions. We also plan to include crustal effects, which are essential for a more accurate treatment of tidal deformability. Additionally, we are interested in investigating non-radial oscillation modes—such as the fundamental (f) and pressure (p) modes—as they serve as key sources of gravitational waves and offer a promising avenue for probing the internal structure of neutron stars through multi-messenger observations.

\section*{Acknowledgement}
This research project is supported by National Research Council of Thailand (NRCT) : (Contact No. N41A670401). CP is supported by Fundamental Fund 2567 of Khon Kaen University. We also thank the referees for their constructive comments and suggestions.

\appendix
\section{Appendices} 
\subsection{The effective energy of nucleon and dark matter} \label{sect:app1}
Recall the equation of motion of nucleon,
\begin{eqnarray}
    \left[ i\gamma^\mu\partial_\mu - g_\omega \gamma^0\omega_0 - g_\rho \gamma^0 \tau_3 b_0 - M^* - g_{aff} i \gamma^5  a_0 \right]\psi &=& 0, \label{eq:A1} 
\end{eqnarray}
we apply $\psi = \psi e^{i k\cdot x-ie(k)t}$, the equation of motion can be expressed as
\begin{eqnarray}
    \left[ i\gamma^\mu\partial_\mu - g_\omega \gamma^0 \omega_0 - g_\rho \gamma^0 \tau_3 b_0 - M^* - g_{aff} i \gamma^5  a_0 \right]\psi &=& 0, \nonumber \\
    \left[ i^2\gamma^i k_i - i^2\gamma^0 e(k)  - g_\omega\gamma^0 \omega_0 - g_\rho \gamma^0 \tau_3 b_0 - M^* - g_{aff} i \gamma^5  a_0 \right]\psi &=& 0.
\end{eqnarray}
Then we multiply the equation with $\gamma^0$, the above equation can be written as
\begin{eqnarray}
    \beta^2\left[e(k) - g_\omega \omega_0 - g_\rho \tau_3 b_0\right]\psi &=& \beta^2\left[ \alpha^i k_i + \beta M^* + g_{aff} i (\gamma^0\gamma^5)  a_0 \right]\psi, \label{eq:A4}
\end{eqnarray}
where 
\begin{eqnarray}
    \alpha^i = \begin{pmatrix}
        0 & \sigma^i \\
        \sigma^i & 0
    \end{pmatrix} \quad \text{and} \quad \beta = \begin{pmatrix}
        I_{2\times2} & 0 \\
        0 & -I_{2\times2}
    \end{pmatrix}.
\end{eqnarray}
We can write equation \ref{eq:A4} as 
\begin{eqnarray}
    \left(e(k) - g_\omega \omega_0 - g_\rho \tau_3  b_0\right)\psi &=& \left(  \begin{pmatrix}
        0 & \sigma^i \\
        \sigma^i & 0
    \end{pmatrix} k_i + 
     \begin{pmatrix}
        I & 0 \\
        0 & -I
    \end{pmatrix}m^* + 
     \begin{pmatrix}
        0 & I \\
        -I & 0
    \end{pmatrix} i g_{aff} a_0 \right)\psi, \nonumber \\
    &=& \begin{pmatrix}
        M^* & \sigma^i k_i + i g_{aff} a_0 \\
        \sigma^i k_i - i g_{aff} a_0 & M^*
    \end{pmatrix} \psi.
\end{eqnarray}
Then multiply by its complex conjugate, we find
\begin{eqnarray}
    \left(e(k) - g_\omega \omega_0 - g_\rho \tau_3 b_0 \right)^2 \psi\bar{\psi} &=& \begin{pmatrix}
        (M^*)^2 + \sigma^i\sigma^j k_i k_j - i^2 g_{aff}^2 a_0^2 & 0\\
        0 & (M^*)^2 + \sigma^i\sigma^j k_i k_j - i^2 g_{aff}^2 a_0^2 
    \end{pmatrix} \psi\bar{\psi}, \nonumber \\
    &=& \left(k^2 + (M^*)^2 +g_{aff}^2 a_0^2 \right)\psi\bar{\psi}.
\end{eqnarray}
We can modify the effective mass of the nucleon as
\begin{eqnarray}
    \widetilde{m}^2 = (m^*)^2 + g_{aff}^2 a_0^2 = (M + g_s\sigma_0)^2 + g_{aff}^2 a_0^2,
\end{eqnarray}
with the energy solution around $g_\omega \omega_0$ as
\begin{eqnarray}
    e^\pm_\psi(k) = g_\omega \omega_0+ g_\rho \tau_3 b_0 \pm \sqrt{\bm{k}^2 + \widetilde{m}^2}. \label{eq:efnew}
\end{eqnarray}

We can repeat the same method for DM. Starting with the equation of motion of DM,
\begin{eqnarray}
    (i\gamma^\mu\partial_\mu - m_\chi -g_{a\chi\chi}i\gamma^5a_0)\chi = 0.
\end{eqnarray}
We can define the modified effective DM mass as 
\begin{eqnarray}
    \widetilde{m}_\chi^2 = m_\chi^2 + g^2_{a\chi\chi} a_0^2,
\end{eqnarray}
with the energy solution
\begin{eqnarray}
    e_\chi^\pm(q) = \pm \sqrt{q^2 + \widetilde{m}_\chi^2}. \label{eq:exnew}
\end{eqnarray}

\bibliographystyle{jhep}
\bibliography{ref}

\end{document}